\documentclass[12pt]{article}
\usepackage{times}
\usepackage{amssymb}

\usepackage[dvips]{graphicx}

\newcommand{\bs}{{\vskip .5cm}}

\begin{document}

\pagestyle{empty}
\begin{center}
{\bf Ordered Morphologies
of Confined Diblock Copolymers}
\footnote{
Submitted to {\it Materials Research Society 
Symposium Proceedings}, {\sl
Dynamics
in Small Confining Systems}, MRS Fall Meeting 2000.}
\end{center}
\noindent {\bf Yoav Tsori and David Andelman}\\ School of Physics
and Astronomy\\  Raymond and Beverly Sackler Faculty of Exact
Sciences\\  Tel Aviv University, 69978 Ramat Aviv, Israel

\raggedright 
\setlength{\parindent}{0.5cm}

\bs
\noindent
{\bf ABSTRACT}
\bs

We investigate the ordered morphologies occurring in
thin-films diblock copolymer. For temperatures above the
order-disorder transition and for an arbitrary
two-dimensional surface pattern, we use
 a Ginzburg-Landau expansion of the free energy
 to obtain a linear response description of the copolymer
melt.
The ordering in the
directions perpendicular and parallel to the surface are
coupled. Three dimensional structures
existing when a melt is confined between two surfaces
are examined.
Below the order-disorder transition we find tilted lamellar
phases in the presence of striped surface fields.

\bs
\noindent
{\bf INTRODUCTION}
\bs

The self-assembly of block copolymers (BCP) has been the subject
of numerous studies \cite{B-F90}-\cite{matsenJCP97}. These
macromolecules are made up of chemically distinct subunits, or
blocks, linked together with a covalent bond. This bonding
inhibits macrophase separation, and leads to formation of
mesophases with typical size ranging from nanometers up to
microns. For di-block copolymers, which are made up of two
partially incompatible blocks (A and B), the phase diagram is
well understood \cite{B-F90}-\cite{M-B96}, and exhibits
disordered, lamellar, hexagonal and cubic micro-phases. The phase
behavior is governed by three parameters: the chain length
$N=N_A+N_B$, the fraction $f=N_A/N$ of the A block, and  the
Flory parameter $\chi$, being inversely proportional to the
temperature.

The presence of a confining surface leads to various
interesting phenomena \cite{Fredrickson87}-\cite{M-RPRL97}.
The surface limits the number of accessible chain
configurations and thus may lead to chain frustration.
Transitions between  perpendicular and parallel lamellar
phases with respect to the confining surfaces  have been
observed  in thin films \cite{M-MPRE96}-\cite{P-WPRE99}. In
addition, the surface may be chemically active, preferring
adsorption of one of the two blocks, and usually
stabilizing the formation of lamellae parallel to the
surface \cite{matsenJCP97,P-BMM97,G-M-B00}.

More complicated behavior occurs when the surface is chemically
heterogeneous; namely, one surface region prefers one block while
other regions prefer the second block. Compared to bulk systems,
new energy and length scales enter the problem adding to its
complexity. Thin-film BCP in presence of chemically patterned
surfaces is of importance in many applications, such as
dielectric mirrors and waveguides \cite{fink98}, anti-reflection
coating for optical surfaces \cite{M-SSCIENCE99} and fabrication
of nanolithographic templates \cite{Chaikin97}.

\newpage

\bs
\noindent
{\bf THE MODEL}
\bs

The copolymer order parameter $\phi$ is defined as
$\phi({\bf r})\equiv\phi_A({\bf r})-f$,  the deviation of
the local A monomer concentration from its average $f$.
Consider a symmetric ($f=1/2$) BCP melt in its disordered
phase (above the bulk ODT temperature) and confined by one
or two flat, chemically patterned surfaces.
The free
energy (in units of the thermal energy $k_BT$) can be
written as
\cite{Leibler80,B-F-SJPII97,F-H87,T-A-S00,N-A-SPRL97}:

\begin{equation}\label{F}
F=\int\left\{\frac12\tau\phi^2+\frac12h\left[\left(\nabla^2+
q_0^2\right)\phi\right]^2
+\frac{u}{4!}\phi^4-\mu\phi\right\}{\rm d}^3{\bf r}
\end{equation} The polymer radius of gyration $R_g$ sets
the periodicity scale $d_0=2\pi/q_0$ via the relation
$q_0\simeq 1.95/R_g$. The chemical potential is $\mu$ and
the other two parameters are $h=1.5\rho c^2R_g^2/q_0^2$ and
$\tau=2\rho N\left(\chi_c-\chi\right)$. The Flory parameter
$\chi$ measures the distance from the ODT point
($\tau=0$),  having the value $\chi_c\simeq 10.49/N$. For
positive $\tau\sim \chi_c-\chi$, the system is in the
disordered phase having $\phi=0$, while for $\tau<0$ (and
$f=1/2$) the lamellar phase has the lowest energy.
Finally, $\rho=1/Na^3$ is the
chain density per unit volume, and $c$ and $u/\rho$ are
constants of order unity \cite{B-F-SJPII97}. This
Ginzburg-Landau expansion of the free energy in powers of
$\phi$ and its derivatives can be justified near the
critical point, where $\tau\ll hq_0^4$ and ordering is
weak. The lamellar phase can approximately be described by a
single $q$-mode there:
$\phi=\phi_q\cos({\bf q_0\cdot r})$.  We note that
 similar types of energy functionals have been used
to describe bulk and surface phenomena
\cite{SH77,AndelmanSC95,mukamelPRE00}, amphiphilic systems
\cite{G-S90,G-ZPRA92}, Langmuir films \cite{A-B-J87} and
magnetic (garnet) films \cite{G-D82}.

We model the chemically heterogeneous surfaces  by a
short-range surface interaction, where the BCP
concentration at  the surface is coupled linearly to a
surface field $\sigma({\bf r_s})$:
\begin{equation}\label{Fs} F_s=\int\sigma({\bf
r_s})\phi({\bf r_s}){\rm d^2}{\bf r_s} \end{equation}
The vector ${\bf r}={\bf r_s}$ defines the position of the
confining surfaces. Preferential adsorption of the A block
is modeled by a $\sigma<0$ surface field, and a constant
$\sigma$ results in a parallel-oriented lamellar layers (a
perpendicular orientation of the chains). Without any
special treatment, the surface tends to prefer one of the
blocks, but by using random copolymers
\cite{L-RPRL96,M-RPRL97} one can reduce this affinity or
even cancel it altogether. A surface with spatially
modulated pattern, $\sigma({\bf r_s}) \ne 0$, induces preferential
adsorption of A and B blocks to different regions of the surface.

\bs
\noindent
{\bf RESULTS AND DISCUSSION}
\bs

We consider first a system of polymer melt in contact with a single
flat surface. The system is assumed to be above the ODT
temperature, in the disordered bulk phase. The results are then
extended to two confining surfaces and to
BCP systems below the ODT.

\bs
\noindent
{\bf Above ODT}
\bs

Above the ODT the bulk phase is disordered, and the free energy,
Eq. (\ref{F}), is convex to second order in the order parameter
$\phi$. Thus, the $\phi^4$ term can be neglected. The melt is
confined to the semi-infinite space $y>0$, bounded by the ${\bf
r_s}=(x,y=0,z)$ surface. The chemical pattern $\sigma({\bf
r_s})=\sigma(x,z)$ can be decomposed in terms of its inplane 
${q}$-modes
$\sigma(x,z)=\sum_{\bf q}\sigma_{\bf
q}\exp\left(i\left(q_xx+q_zz\right)\right)$, where ${\bf
q}\equiv(q_x,q_z)$, and $\sigma_{\bf q}$ is the mode amplitude.
Similarly, the order parameter is $\phi(x,y,z)=\sum_{\bf
q}\phi_{\bf q}(y)\exp\left(i\left(q_xx+q_zz\right)\right)$.
Substituting $\phi$ in Eq.~(\ref{F}), and applying a variational
principle with respect to $\phi_{\bf q}$, results in a linear
fourth-order differential equation \cite{TAMM01,TAEPL01}:
\begin{eqnarray}\label{EL-fq}
\left(\tau/h+\left(q^2-q_0^2\right)^2\right)\phi_{\bf q}(y)
+2(q_0^2-q^2)\phi_{\bf q}^{\prime\prime}(y)+\phi_{\bf
q}^{\prime\prime\prime\prime}(y)=0 \end{eqnarray}
In the semi-infinite geometry, $y>0$, the solution to
Eq.~(\ref{EL-fq}) has an exponential form $\phi_{\bf
q}(y)=A_{\bf q}\exp(-k_{\bf q}y)+B_{\bf q}\exp(-k^*_{\bf
q}y)\label{fq}$, where $k_{\bf q}$ is given by
\begin{eqnarray}
k_{\bf q}^2
&=&q^2-q_0^2+i\sqrt{\tau/h}\nonumber\\
&=&q^2-q_0^2+i\alpha\left(N\chi_c-N\chi\right)^{1/2}\label{kq}
\end{eqnarray}
with $\alpha\simeq 0.59 q_0^2$. The values of $\xi_q=1/{\rm
Re}(k_{\bf q})$ and $\lambda_q=1/{\rm Im}(k_{\bf q})$ correspond
to the exponential decay and oscillation lengths of the surface
${q}$-modes, respectively. For fixed $\chi$, $\xi_q$ decreases
and $\lambda_q$ increases with increasing $q$
\cite{P-Muthu97-98,TAMM01,TAEPL01}. Close to the ODT (but within
the range of validity of the model), and for ${q}$-modes such
that $q>q_0$ we find finite $\xi_q$ and
$\lambda_q\sim(\chi_c-\chi)^{-1/2}$.  The propagation of the
surface imprint (pattern) of $q$-modes with $q<q_0$ into the bulk
can persist to long distances, in contrast to surface patterns
with $q>q_0$ which decays off close to the surface.  This is seen
by noting that $q$-modes in the band $0<q<q_0$ are equally
``active'', i.e., these modes decay to zero very slowly in the
vicinity of the ODT as $y\rightarrow\infty$:
$\xi_q\sim(\chi_c-\chi)^{-1/2}$ and $\lambda_q$ is finite.

The boundary conditions for $\phi_{\bf q}$ at $y=0$  are
\begin{eqnarray}\label{bcs} \phi_{\bf
q}^{\prime\prime}(0)+\left(q_0^2-q^2\right) \phi_{\bf
q}(0)&=&0\\ \sigma_{\bf
q}/h+\left(q_0^2-q^2\right)\phi_{\bf
q}^{\prime}(0)+\phi_{\bf q}^{\prime\prime\prime}(0)
&=&0\nonumber \end{eqnarray}
The amplitude $A_q$ is found to be $A_q=-\sigma_q\left(2{\rm
Im}(k_q)\sqrt{\tau h}\right)^{-1}$. Thus, the copolymer response
diverges upon approaching the critical point as
$\left(N\chi_c-N\chi\right)^{-1/2}$.

\begin{figure}[h]
\begin{center}
\includegraphics[scale=0.55]{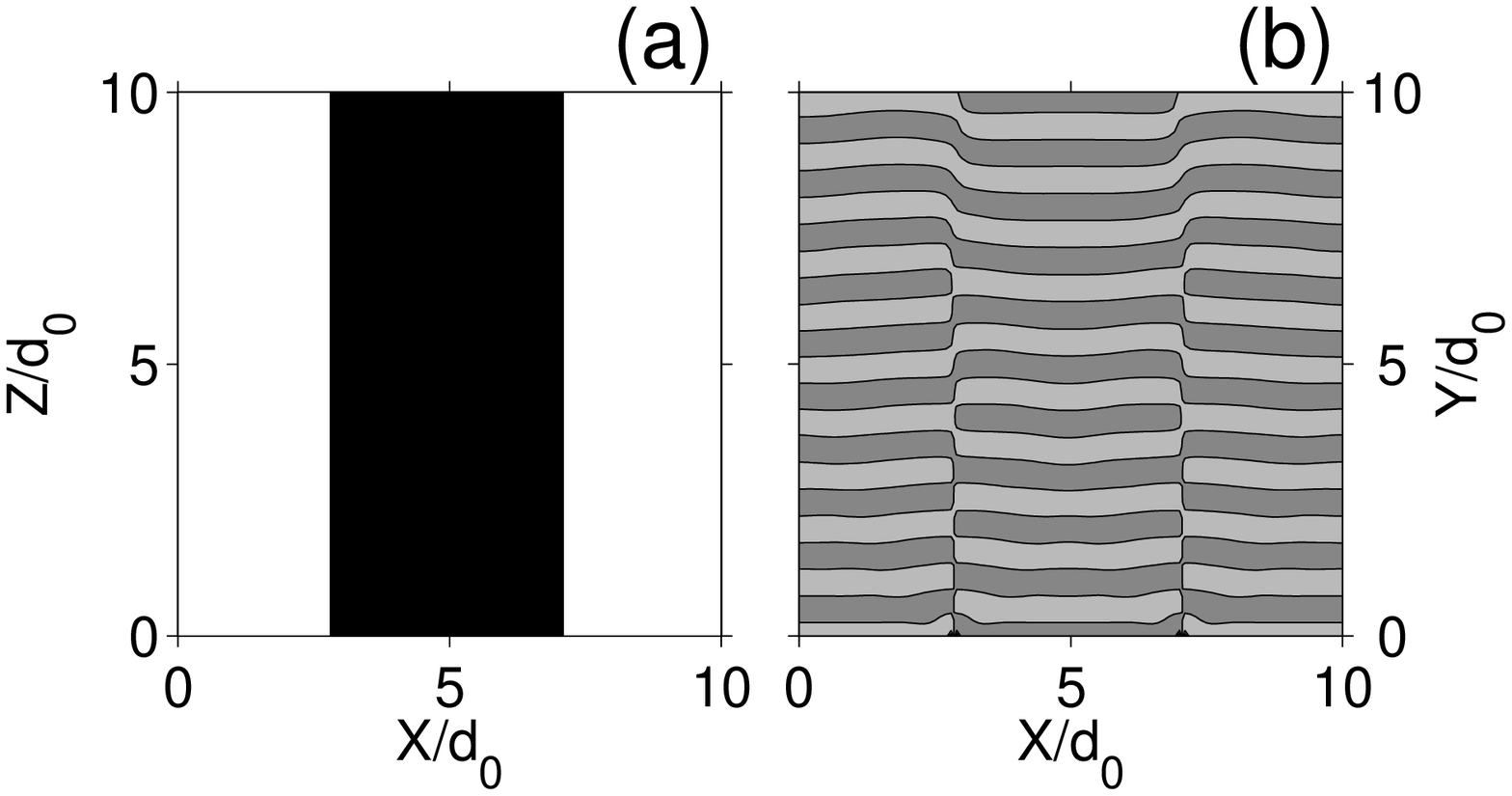}
\end{center}
\noindent \footnotesize{{\bf Figure 1.} \it A BCP melt confined
by one patterned surface having a central stripe of width
$5d_0$.  In (a) the surface located at $y=0$ has a localized
``surface disturbance'' of $\sigma=1/2$ inside the black stripe,
favoring the adsorption of the B monomers, and $\sigma=-1/2$
outside of the stripe (favoring A monomers). The morphology  in
the $x$-$y$ plane is shown in (b), where farther away from the
stripe (large $y\gg d_0$) the lamellae have a decaying order,
while close to the surface ($y\lesssim d_0$) the lamellae are
distorted to optimize the interfacial energy. The system is above
the ODT with $\chi N=10$.  Lengths are scaled by $d_0$, the
lamellar  periodicity. B-rich regions are mapped to black shades
and A-rich to white. In all plots we set the monomer length as
$a=1$, and choose in Eq. (\ref{F}) $c=u/\rho=1$, $R_g^2=\frac16
Na^2$ and $N=1000$ to give $\alpha\simeq0.59q_0^2$. }
\end{figure}
Although our analysis allows for arbitrary
surface patterns, let us consider first the simple
case of a surface
attractive for the A block, but which has a single localized
stripe, or surface ``disturbance'', preferentially
attracting the B
block (see Fig. 1~(a)).
\begin{figure}[!h]
\begin{center}
\includegraphics[scale=0.55]{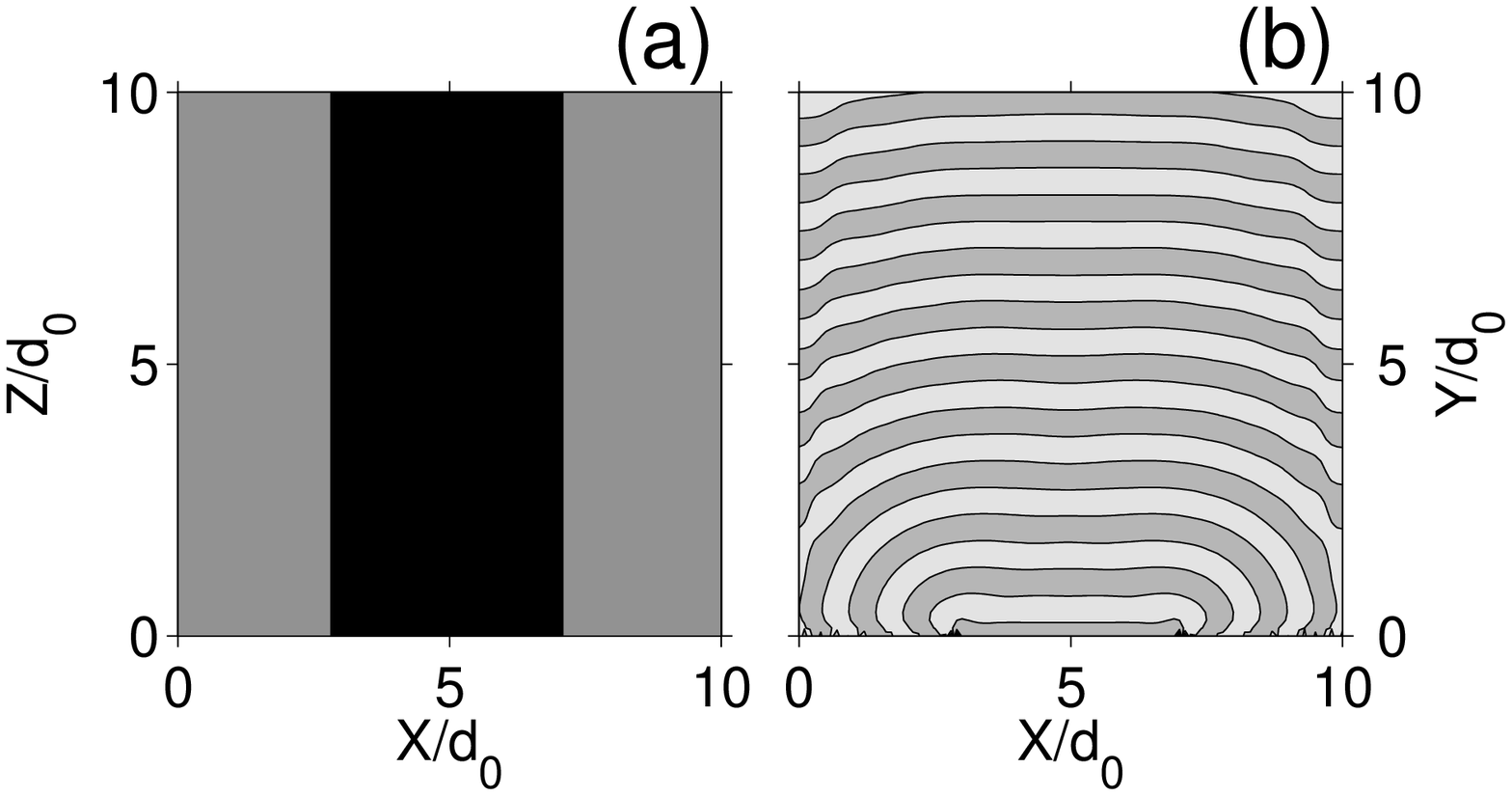}
\end{center}
\footnotesize{{\bf Figure 2.}  \it Same as in
Fig. 1, but here the surface area outside of the stripe is
neutral to polymer adsorption, $\sigma=0$. The B monomers
are found close to the stripe, inducing a curved lamellar
structure that decays away from the surface.}
\end{figure}
Far from this stripe A-blocks are adsorbed on the surface and
the copolymer concentration profile has decaying lamellar
order as was explained above. However, a non-trivial BCP
morphology appears close to the stripe, reflecting the
adsorption of B monomers. This behavior is indeed seen in
Fig.~1 (b), where the grey scale
is such that A-rich regions are white, while
B-rich are black. A somewhat different situation is shown
in Fig.~2 (a), where the black stripe is still preferential to
the B monomers, but the rest of the surface (in grey) is neutral,
$\sigma(x,z)=0$.  In this case no lamellar ordering
parallel to the surface is expected. In (b) curved lamellae
appear around the surface disturbance, optimizing
interfacial energy. These curved lamellae are exponentially
damped both in directions parallel and perpendicular to the
surface. Clearly, the ordering in the perpendicular and
lateral directions are coupled.

A different scenario is presented in Fig.~3, where the surface
has a sinusoidal variation in its affinity for the monomer type:
$\sigma(x)=\sigma_0+\sigma_q\cos(qx)$, with periodicity $d=2\pi/q$
and average attraction $\langle\sigma\rangle=\sigma_0>0$ for the B
monomers. The modulated surface field of amplitude $\sigma_q$
induces lateral order, with A- and B-rich regions near the
surface. Near the top of the figure ($y\gg d_0$), only
lamellar-like ordering is seen,  because the $\sigma_q$ ($q>0$)
term decays faster than the $\sigma_0$ ($q=0$) term.

Using this formulation, any two-dimensional chemical pattern
$\sigma(x,z)$ can be modeled. For surface feature of size larger
than $d_0$, the characteristic copolymer length, the chemical
surface pattern can propagate via the BCP melt into the bulk.
\begin{figure}[!h]
\begin{center}
\includegraphics[scale=0.4]{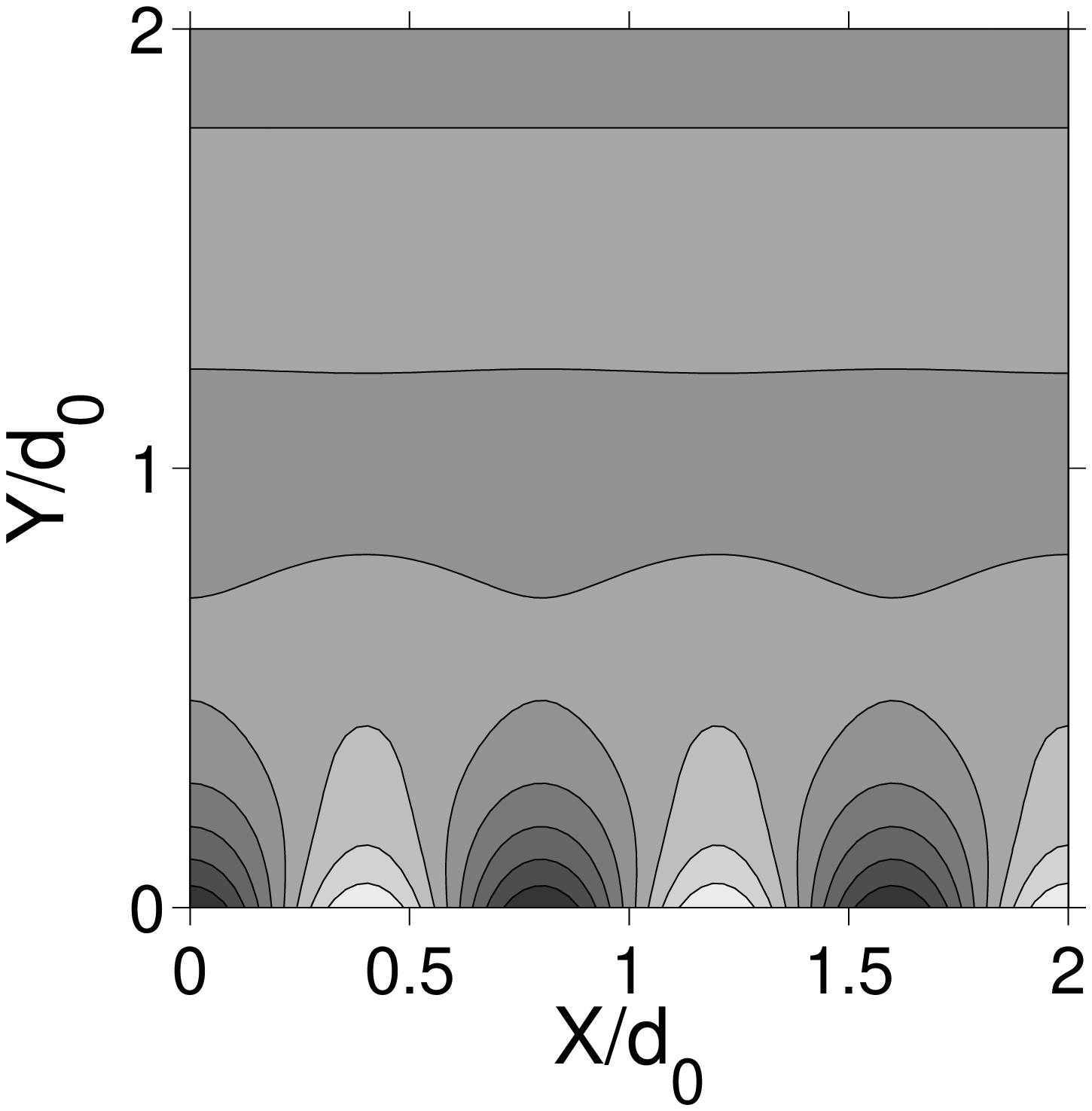}
\end{center}
\noindent
\footnotesize{{\bf Figure 3.} {\it  Copolymer morphology
for a melt confined by one surface  at $y=0$ having a
chemical affinity of the form $\sigma(x)=
\sigma_0+\sigma_q\cos(q_xx)$, with $q_x=\frac54q_0$ and
$\sigma_0= \sigma_q=0.1$. The lamellar-like order due to
the $\sigma_0$ term persists farther away from the
wall than the lateral order due to the $\sigma_q$ term.
The Flory parameter is $\chi N=10.4$.}}
\end{figure}
To demonstrate this we take in Fig.~4 (a) a surface pattern in
the shape of the letters `MRS'. Inside the letters $\sigma>0$ (B
monomers are attracted), while for the rest of the surface
$\sigma=0$ (neutral). The resulting patterns in the $x$-$z$
planes are shown for $y/d_0=0.5$, $2$ and $3.5$ in (b), (c) and
(d), respectively. An overall blurring of the image is seen as
the distance from the surface is increased. The fine details
(e.g. sharp corners of the letters) disappear first as a
consequence of the fast decay of high surface $q$-modes.
\begin{figure}[h]
\begin{center}
\includegraphics[scale=0.5]{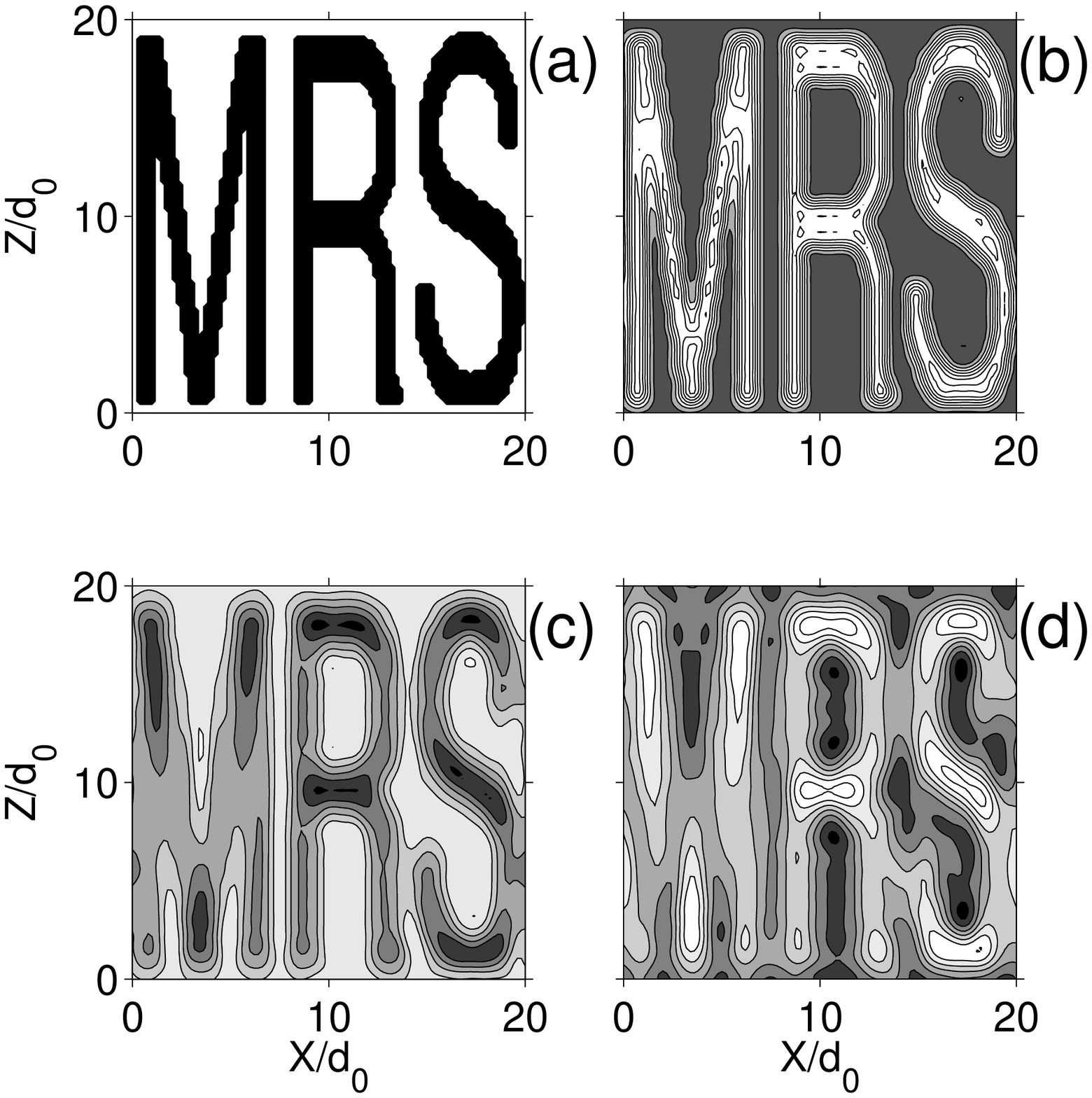}
\end{center}
\noindent \footnotesize{{\bf Figure 4.} \it Propagation of
surface order into the bulk BCP melt. (a) The surface at $y=0$ is
taken to have a pattern in the shape of the letters `MRS'. Inside
the letters ($\sigma=1$) the field attracts the B monomers, while
the rest of the surface is neutral ($\sigma=0$).  In (b), (c) and
(d) the morphology is calculated for $x$-$z$ planes located at
increasing distances from the surface, $y/d_0=0.5$, $2$ and $3.5$,
respectively. Note that in (b) and (d) black (white) shades can
be mapped into white (black) shades in (c), because their
$y$-spacing is a half-integer number of lamellae. The Flory
parameter is $\chi N=9.5$.}
\end{figure}
The second point to notice is the A$\leftrightarrow$B interchange
of monomers that occurs for surfaces separated roughly by a
distance of $(n+\frac12)d_0$, for integer $n$. This monomer
interchange mimics the formation of lamellae in the bulk,
although the temperature here is above the ODT. In addition, the
appearance of lateral order is clearly seen, as lamellae form
parallel to the edges of the letters.

\begin{figure}[!h]
\begin{center}
\includegraphics[scale=0.75]{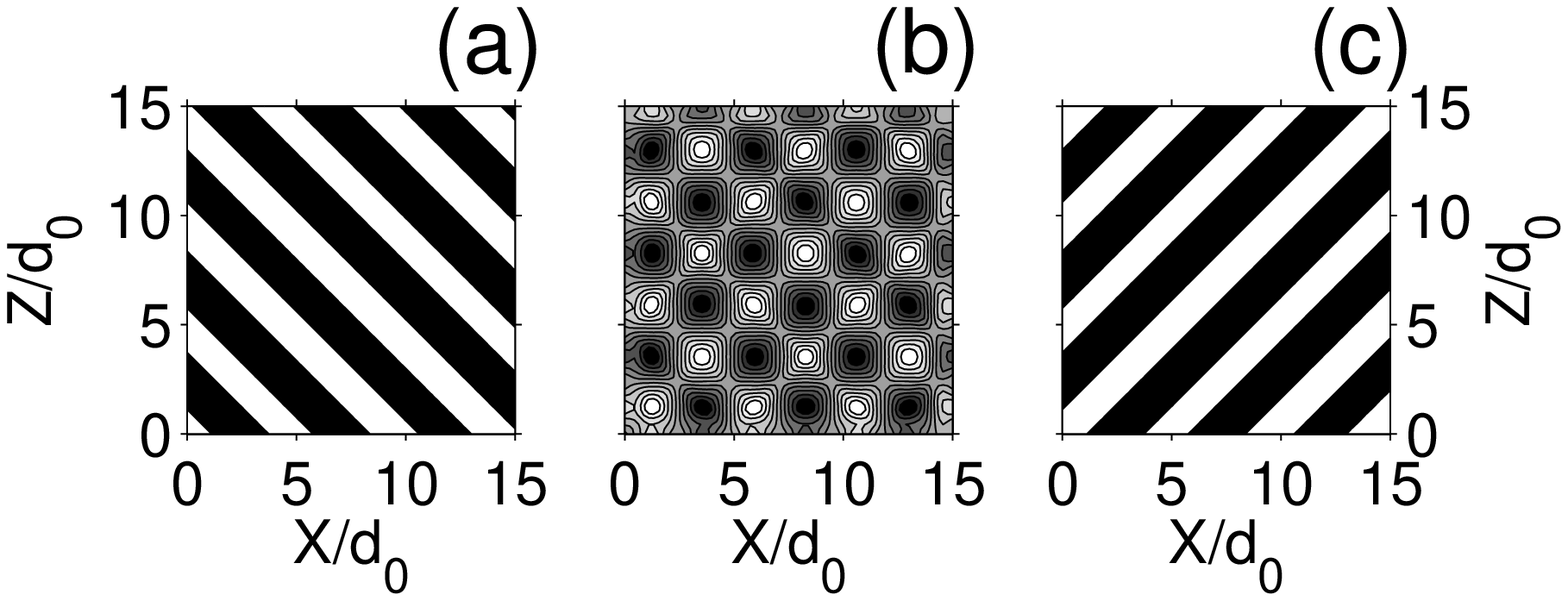}
\end{center}
\noindent \footnotesize{{\bf Figure 5.} \it Creation of three
dimensional structure in a  thin BCP film as a result of two
one-dimensional surface patterns. (a) and (c) are stripe surfaces
located at $y=-L=-d_0$ and $y=L=d_0$, respectively. The mid-plane
($y=0$) ordering is shown in (b) and is created by a
superposition of the two surface fields. $\chi N=9$. }
\end{figure}

We briefly mention the case where a BCP melt is confined between
two flat parallel surfaces located at $y=\pm L$.
The calculation of the response functions $\{\phi_{\bf q}\}$ can
easily be generalized to handle two confining surfaces  by
including the appropriate boundary conditions in Eqs.
(\ref{bcs}). If the distance $2L$ is very large, the copolymer
orderings induced by the two surfaces are not coupled, and the
middle of the film ($|y|\ll L$) is disordered, $\phi\approx 0$.
Decreasing the film thickness to a distance comparable to the
copolymer correlation length results in an overlap of the
two surface fields.

Complex three dimensional morphologies can
also be achieved by using only one dimensional surface patterns,
if the two patterns are rotated with respect to each other.
Such an example is shown in Fig.~5, for two surfaces at
$y=\pm L = \pm d_0$ with perpendicularly oriented stripes. The surface
patterns are shown in (a) and (c), while the mid-plane
checkerboard morphology ($y=0$) is shown  in (b). In the
next subsection we will show results for the more
complicated situation of BCP melt below the ODT
temperature, where the bulk phase is lamellar.

\newpage
\bs
\noindent
{\bf Below ODT}
\bs

The prevailing bulk phase below the ODT has an inherent lamellar
ordering. This order interferes strongly with the surface induced
order, and it is not a simple task to obtain order-parameter
expressions as a function of arbitrary surface pattern.
Mathematically, the difficulty lies in the fact that the $\phi^4$
term in the free energy cannot be neglected. As in the case above
the ODT, we expand the free energy to second order around the
bulk phase, only this time the bulk phase is lamellar. As we will
see, this approach has also been used to describe defects in bulk
phases such as chevron and omega-shaped tilt grain boundaries
\cite{T-A-S00,N-A-SPRL97}. We consider first a BCP melt confined
by one stripe surface whose one dimensional pattern is of the
form:
\begin{equation}
\sigma(x,z)=\sigma_q\cos(q_xx)
\end{equation}
The surface periodicity $d_x=2\pi/q_x$ is assumed to be larger
than the bulk lamellar spacing $d_0=2\pi/q_0$, $d_x>d_0$
throughout the analysis. The system reduces its interfacial energy
by trying to follow the surface modulations, and an overall
 tilt of the lamellae follows regardless of the fine details of chain
conformations near the surface. The tilt angle with respect
to the surface is defined as $\theta=\arcsin(d_0/d_x)$
\cite{P-Muthu97-98}. Consequently, we use the single
$q$-mode approximation to describe  the bulk phase $\phi_b$
\begin{equation}
\phi_b(x,y)=-\phi_q\cos(q_xx+q_yy)
\end{equation}
where $q_y\equiv q_0\sin(\theta)$ and $q_x=q_0\cos(\theta)$. All
surface effects are contained in the correction to the order
parameter: $\delta\phi({\bf r})=\phi({\bf r})-\phi_b({\bf r})$.
We choose the in-phase one harmonic form for $\delta\phi$
\begin{equation}
\delta\phi(x,y)=g(y)\cos(q_xx)
\end{equation}
with a $y$-dependent amplitude $g(y)$. This correction
$\delta\phi$ is expected to vanish far from the surface,
$\lim_{y\rightarrow\infty}\delta\phi=0$, recovering the bulk
phase. The free energy in Eq. (\ref{F}) is expanded to second
order in the small correction $\delta\phi$, $\phi_b\gg
\delta\phi$. We integrate out the $x$ dependence and  retain only
the $y$ dependence. Then, use of a variational principle with
respect to the function $g(y)$ yields a linear differential
equation:
\begin{equation}\label{g_gov} \left[ A-C\cos(2q_yy)
\right]g+Bg^{\prime\prime} +g^{\prime\prime\prime\prime}=0
\end{equation}
where $A$, $B$, and $C$ are parameters
depending on the tilt angle $\theta$ as well as on the
temperature \cite{T-A-S00}. This equation is similar to the
Mathieu equation describing an electron under the influence
of a periodic  one-dimensional potential, and has a
solution in the form of a Bloch wave function.

\begin{figure}[h]
\begin{center}
\includegraphics[scale=0.75]{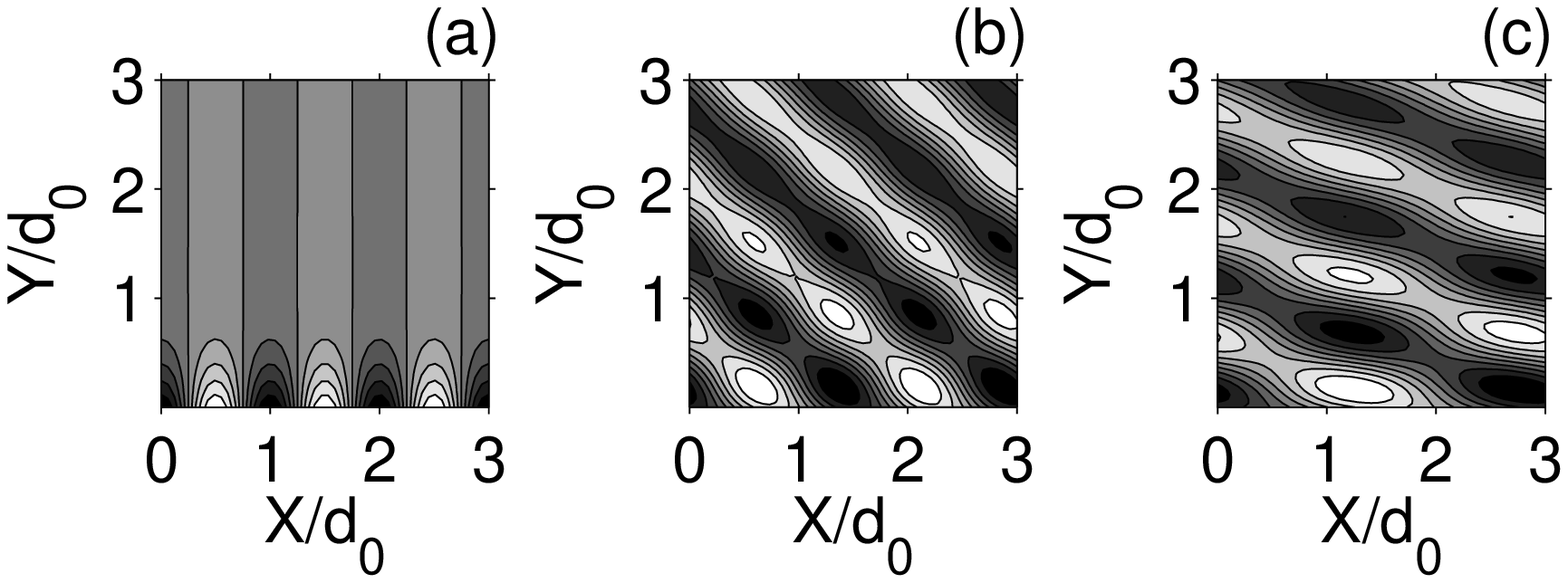}
\end{center}
\noindent \footnotesize{{\bf Figure 6.} \it BCP melt confined by
one stripe surface of periodicity  $d_x>d_0$ below the ODT. The
surface stripes are described by $\sigma(x)=\sigma_q\cos(q_xx)$.
Tilted lamellae with respect to the surface  at $y=0$ are formed
and adjust to the surface imposed periodicity.  Far from the
surface the lamellae relax to their   undistorted $d_0$ spacing.
The mismatch ratio $d_x/d_0$ is $1$ in (a)  yielding no tilt,
$3/2$ in (b) and $3$ in (c).  The Flory parameter is taken as
$\chi N=10.6$ and $\sigma_q=0.08$. }
\end{figure}
The main effect of the surface stripes on the melt is to
introduce a tilt of the lamellae. This effect is seen in Fig.~6,
where the surface periodicity $d_x$ (in units of $d_0$) is chosen
to be $1$ in (a), and induces a perpendicular ordering. Larger
ratios of $d_x/d_0$ cause a tilt ordering. In (b) $d_x/d_0=3/2$
and in (c) $d_x/d_0=3$, and the deviation from a perfect lamellar
shape can be seen near the surface.

\begin{figure}[h]
\begin{center}
  \includegraphics[scale=0.3]{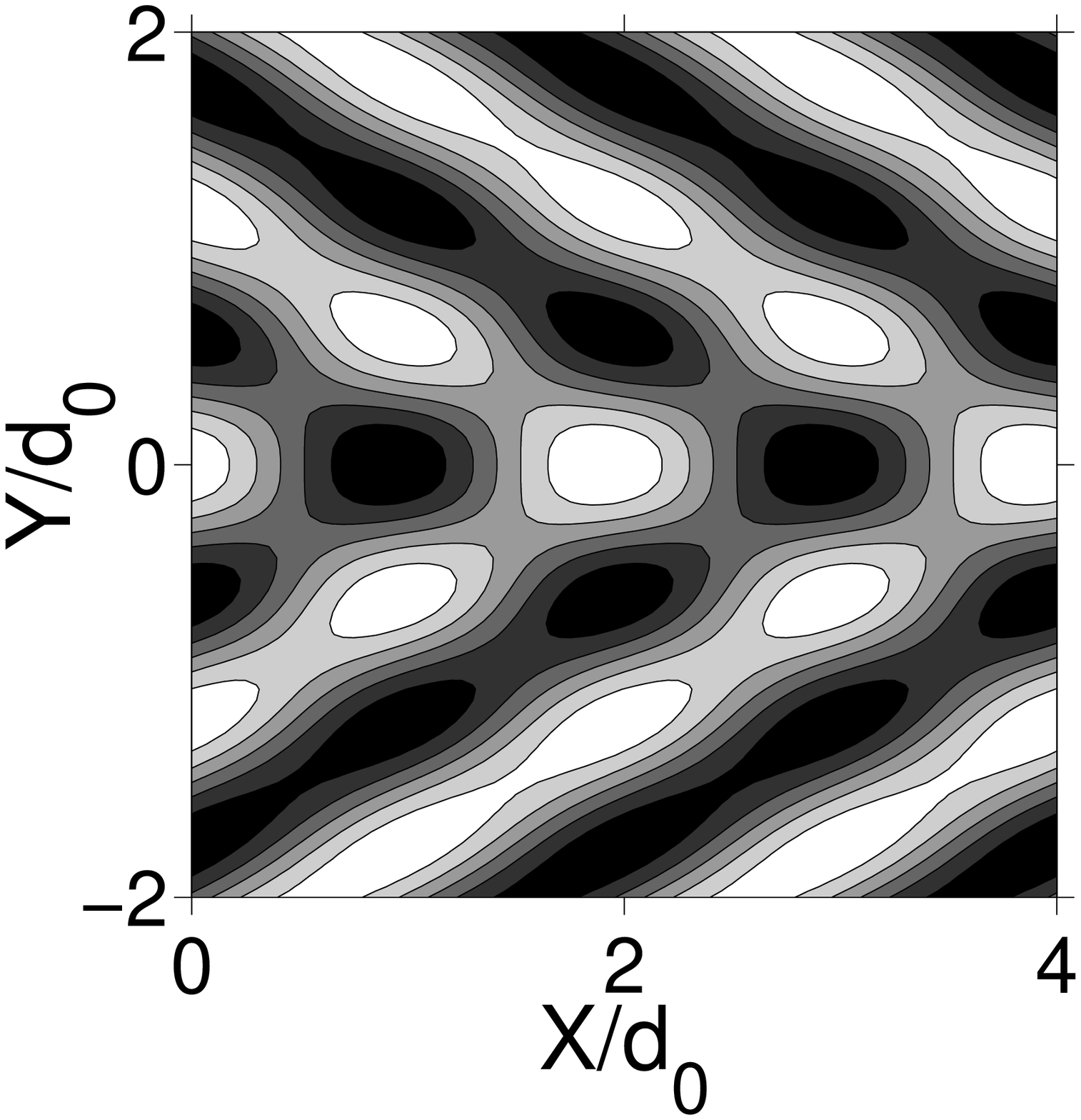}
\end{center}
\noindent
\footnotesize{{\bf Figure 7.} \it Symmetric tilt grain
boundary between two bulk lamellar grains.
The angle between lamellae is determined by external
boundary conditions at $y\rightarrow\infty$ and is chosen here
to be $\varphi=120^\circ$}.
The morphology
is invariant in the $z$ direction and $\chi  N=11$.

\end{figure}

\bs
\noindent
{\bf Tilt Grain Boundaries}
\bs

Our approach can be used to describe defects in bulk systems as
well. A tilt grain boundary forms when two lamellar bulk grains
meet with a tilt angle $\varphi$ between the lamellae normals.
For symmetric tilt grain boundaries, the plane of symmetry is
analogous to the patterned surface in a  thin BCP film, and
similar form of a correction field $\delta\phi$ can be used, see
Fig.~7. The upper half plane $y>0$ thus corresponds to Fig.~6 (b)
and (c).

The correction is small in the so-called chevron region (small
tilt angle $\varphi$), but becomes important for larger tilt
angles \cite{T-A-S00} where the grain boundary has the form of the
letter Omega. Close to the ODT point the modulations at the
interface become long range and persist into the bulk up to large
distances. One of the differences between tilt grain boundaries
and tilt induced by surface field is that in the former case there
are no real surface fields. The tilt angle between adjacent
grains is determined by constraints imposed far from the $y=0$
interface.

\newpage

\bs
\noindent
{\bf CONCLUSIONS}
\bs

An analytical expansion of the free energy, Eq. (\ref{F}), can be
used to obtain order parameter expressions for a BCP melt confined
by one or two flat surfaces whose pattern has arbitrary (two
dimensional) shape. The thin film morphology can also have a
complex three dimensional form even if the two surfaces have one
dimensional patterns with different orientations  with respect to
each other, see Fig.~5. A BCP melt close to and above the ODT
point shows decaying oscillations of the local concentration
towards the bulk disorder value.  Our analysis shows that close
to the ODT large surface spatial features can be transferred into
the  BCP bulk far from the surface, while small surface details
are greatly damped.

Below the ODT, confinement of the melt by patterned surfaces with
periodicity $d_x>d_0$ leads to a formation of a surface layer
characterized by lateral periodicity of $d_x$, and the lamellae
relax to their natural periodicity $d_0$ farther from the surface.
The proposed mechanism is a tilting of the lamellae. We show that
the resulting pattern have similar characteristics to bulk domain
walls (tilt grain boundaries).

\bs
\noindent
{\bf ACKNOWLEDGMENTS}
\bs

We would like to thank M. Muthukumar, R. Netz, G. Reiter, T.
Russell, M. Schick and U. Steiner for useful comments and
discussions. Partial support from the U.S.-Israel Binational
Foundation (B.S.F.) under grant No. 98-00429 and the Israel
Science Foundation funded by the Israel Academy of Sciences and
Humanities --- centers of excellence program is gratefully
acknowledged.

\bs

\end{document}